\documentclass[10pt,letterpaper]{article}
\usepackage{opex3}
\usepackage{cite}
\usepackage{units}

\newcommand{\be}{\begin{equation}}
\newcommand{\ee}{\end{equation}}

\begin{document}
\title{An optimized photon pair source for quantum circuits}
\author{Georg Harder,$^{1*}$ Vahid Ansari,$^1$ Benjamin Brecht,$^1$ Thomas Dirmeier,$^{2,3}$ Christoph Marquardt,$^{2,3}$ and Christine Silberhorn$^{1,2}$}
\address{$^1$Applied Physics, University of Paderborn,  \\ Warburger Stra\ss e 100, 33098 Paderborn, Germany \\
$^2$Max Planck Institute for the Science of Light, \\Guenther-Scharowsky-Str. 1/building 24, 91058 Erlangen, Germany \\
$^3$Institute of Optics, Information and Photonics, University of Erlangen-Nuremberg, \\ Staudtstra\ss e 7/B2, 91058 Erlangen, Germany
}
\email{$^*$georg.harder@uni-paderborn.de}

\begin{abstract}
We implement an ultrafast pulsed type-II parametric down conversion source in a periodically poled KTP waveguide at telecommunication wavelengths with almost identical properties between signal and idler. As such, our source resembles closely a pure, genuine single mode photon pair source with indistinguishable modes. We measure the joint spectral intensity distribution and second order correlation functions of the marginal beams and find with both methods very low effective mode numbers corresponding to a Schmidt number below $1.16$. We further demonstrate the indistinguishability as well as the purity of signal and idler photons by Hong-Ou-Mandel interferences between signal and idler and between signal/idler and a coherent field, respectively. Without using narrowband spectral filtering, we achieve a visibility for the interference between signal and idler of $\unit[94.8]{\%}$ and determine a purity of more than $\unit[80]{\%}$ for the heralded single photon states. Moreover, we measure raw heralding efficiencies of $\unit[20.5]{\%}$ and $\unit[15.5]{\%}$ for the signal and idler beams corresponding to detector-loss corrected values of $\unit[80]{\%}$ and $\unit[70]{\%}$. \end{abstract}



\section{Introduction}
Complex photonic systems for quantum communication or quantum simulation applications rely on the interference between different quantum states of light at a beam splitter \cite{knill_scheme_2001, ralph_quantum_2006, spring_boson_2012, broome_photonic_2012}. The indistinguishability and purity of these states is crucial for the performance of such applications as dissimilarities and mixedness reduce the visibility and hence the quantum character of the interference. Quantum state sources for these applications should therefore not only meet demands of brightness and efficiency, but also generate pure and indistinguishable states.

Parametric down-conversion (PDC) is a versatile source of quantum states which can be used either as direct input states or for heralding single photons \cite{uren_efficient_2004, pittman_heralding_2005, lvovsky_quantum_2001, laiho_producing_2009} as well as other non-Gaussian states \cite{zavatta_quantum--classical_2004, wakui_photon_2007, bimbard_quantum-optical_2010}. However, generic PDC sources feature spectral correlations between signal and idler \cite{grice_eliminating_2001}. When single photons are heralded from these sources, the correlations introduce mixedness in the heralded states \cite{uren_generation_2005, rohde_spectral_2007}. Similarly, when interfering states from independent sources, spectral and spatial correlations decrease the visibility of the interference. It is therefore desirable to have separable PDC sources with no correlations between signal and idler at ones disposal. This implies that it must be possible to define for each photon an ultrafast wavepacket, which also occupies exactly one single spatial mode.

The most simple and widely used approach to achieve spectral decorrelation is to use spectral filtering \cite{uren_efficient_2004, pittman_heralding_2005, lvovsky_quantum_2001, laiho_producing_2009, zavatta_quantum--classical_2004, bimbard_quantum-optical_2010, Pittman_violation_2003, tapster_photon_1998}. However, perfect decorrelation can only be achieved in the limit of a small bandwidth, reducing the brightness of the source significantly \cite{branczyk_optimized_2010}. This tradeoff can be avoided by tailoring the dispersion properties of the nonlinear medium as well as the spectrum of the pump beam in such a way that the PDC process produces decorrelated states directly \cite{grice_eliminating_2001, uren_generation_2005}. This approach has first been realized in bulk PDC at wavelengths around \unit[800]{nm} \cite{mosley_heralded_2008}.  For telecom wavelengths, a decorrelated source in a waveguide structure was realized in periodically poled potassium titanyl phosphate (ppKTP) \cite{eckstein_highly_2011}. Alternatively to PDC, four-wave mixing in microstructured fibers \cite{cohen_tailored_2009} or silica waveguides \cite{spring_chip_2013} can also be used to produced decorrelated photon pair states.
The additional condition of indistinguishability between signal and idler has been realized in bulk KTP at telecom wavelengths \cite{gerrits_generation_2011} and in bulk BBO at around \unit[800]{nm} \cite{tanida_highly_2012}. 

In this paper we report on an integrated source that combines decorrelation, indistinguishability and telecom wavelengths without relying on narrowband filtering. Being a waveguide source, it is efficient, exhibits decorrelated spatial Gaussian mode profiles and thus features high coupling efficiencies into single mode fibers allowing for easy integration with existing telecommunication networks. 

Our source is also capable of producing bright two mode squeezed states with a mean photon number well above one. However, in this paper we mainly focus on the characterization as a photon pair source, i.e. we work in the low power regime with a mean photon number far below one.

\section{Theory}
\label{sec:theory}
\subsection{Decorrelated parametric down-conversion}
The spectral properties of a PDC process in waveguides are governed by the energy and momentum conservation laws
\be \omega_s + \omega_i = \omega_p, \qquad k_s + k_i + \frac{2\pi}{\Lambda} = k_p, \ee
where $\omega$ is the frequency of the signal, idler or pump photons, respectively, $k$ their propagation constant and $\Lambda$ the period of a periodic poling of the waveguide.  Energy and momentum are linked by the dispersion relation $\omega(k)$ of the waveguide, mainly influenced by the material properties. For short pump pulses, a distribution of frequencies has to be taken into account, the pump function $\alpha(\omega_p)$. Similarly, for waveguides of finite lengths, the momentum conservation is described by a phasematching function $\phi(\omega_s,\omega_i)$. The full process is then described by the joint spectral amplitude (JSA)
\be f(\omega_s, \omega_i) = \phi(\omega_s, \omega_i)\alpha(\omega_s, \omega_i) \ee
and the Hamiltonian is given by $H = \zeta \int \textrm{d}\omega_s\textrm{d}\omega_i f(\omega_s, \omega_i) a^\dag_s(\omega_s)a^\dag_i(\omega_i) + \textrm{h.c.}$,  where $a^\dag(\omega)$ is the standard creation operator at frequency $\omega$. The amount of spectral correlations can be quantified by the Schmidt number $K = 1/\sum_k |c_k|^4$, where the coefficients $c_k$ result from the Schmidt decomposition of the JSA: $f(\omega_s, \omega_i) = \sum_k c_k \phi_k(\omega_s)\psi_k(\omega_i)$, with orthonormal functions $\phi_k$ and $\psi_k$ \cite{law_continuous_2000}. In particular for $K=1$, signal and idler are decorrelated and the state is separable. 

\begin{figure}[ht]
\begin{center}
\includegraphics*{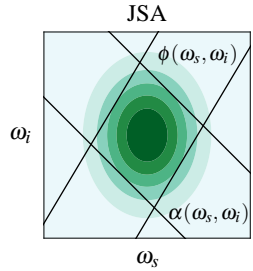}
\caption{Separable joint spectral amplitude resulting from a phasematching function $\phi$ with a positive slope multiplied by a pump function $\alpha$ of appropriate width.}
\label{fig:theory_JSA}
\end{center}
\end{figure}

Most PDC sources exhibit a negatively correlated phasematching function $\phi(\omega_s,\omega_i)$. Since the pump function $\alpha(\omega_s,\omega_i)$ is always oriented exactly at $-45^\circ$ due to energy conservation, spectral correlations cannot be avoided in these sources. With a positively oriented phasematching, however, spectral decorrelation is possible. By dispersion engineering, i.e. choosing the right material, and adjustment of the pump width, spectral decorrelation can be achieved \cite{uren_generation_2005}.  KTP waveguides have a phasematching correlation angle of approximately $59^\circ$ \cite{eckstein_mastering_2012} at telecom wavelengths around $\unit[1550]{nm}$. Assuming a Gaussian pump spectrum and a Gaussian phasematching function, which is a good approximation for realistic, slightly inhomogeneous waveguides, the joint spectral amplitude becomes perfectly separable with a shape that is close to round, see Fig. \ref{fig:theory_JSA}. 

\subsection{Source characterization}
\paragraph{Hong Ou Mandel interference between signal and idler}
\label{sec:HOM_interference}
The interference of twin photons at a beamsplitter followed by a coincidence measurement with single photon detectors is a well established method to deduce information about a biphoton quantum state. First demonstrated by Hong, Ou and Mandel \cite{hong_measurement_1987} to measure the duration of photon wavepackets, it can be used as a measure of indistinguishability \cite{pittman_heralding_2005, mosley_conditional_2008, sun_indistinguishability_2009} and even for reconstruction of the spectral properties of a state \cite{avenhaus_experimental_2009, wasilewski_spectral_2007, kolenderski_derivation_2009}. The Hong-Ou-Mandel (HOM) interference is unaffected by loss and has been widely applied for the characterization of single photon states from PDC. 

\begin{figure}[ht]
\begin{center}
\includegraphics*{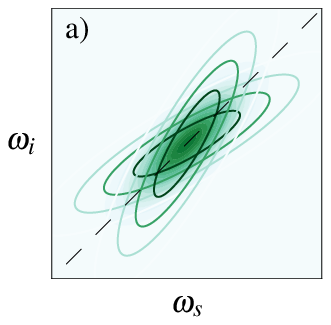} \quad \includegraphics*{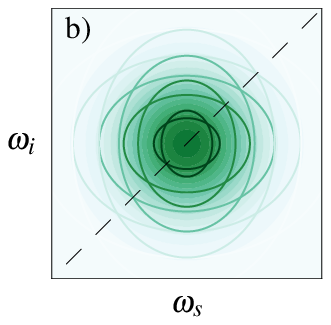} \quad \includegraphics*{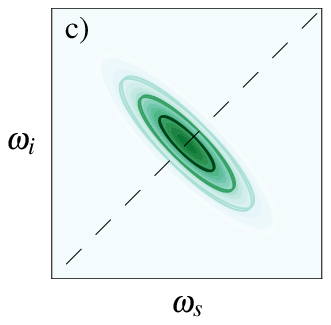}
\caption{The visibility of the HOM interference is proportional to the overlap (shaded area) of the JSA with itself under the exchange of signal and idler (contour plots). a) Positive correlation, $K=2$ and visibility $70\%$. b) Decorrelation, $K=1$ and visibility $96.7\%$. c) Negative correlation, $K=2$ and visibility $99.8\%$. Panel b) corresponds to the situation presented in this paper.}
\label{fig:theory_HOM}
\end{center}
\end{figure}

Assuming that signal and idler originate from a perfect, weakly pumped PDC source with a not necessarily decorrelated JSA $f(\omega_s,\omega_i)$, the probability of a coincidence click after the interference at a $50/50$ beam splitter for a time delay $\tau$ between signal and idler is given by
\be p(\tau) = \frac{1}{2} -\frac{1}{2} \textrm{Re}  \int \textrm{d}\omega_s \textrm{d}\omega_i f^*(\omega_s, \omega_i) f(\omega_i,\omega_s) e^{-i\tau(\omega_s-\omega_i)}. \ee
For $\tau=0$, the integral term is the overlap between the JSA and its mirrored counterpart along the $45^\circ$ line. A HOM measurement hence probes the symmetry of the state under the exchange of signal and idler, as illustrated in Fig. \ref{fig:theory_HOM}. 

\paragraph{Hong Ou Mandel interference with a reference field}
For the phasematching correlation angle of our source, high visibilities in a HOM interference between signal and idler can be obtained in the decorrelated as well as in the negatively correlated case. Our goal, however, is to produce decorrelated states as in the central plot of Fig. \ref{fig:theory_HOM}. To distinguish between the decorrelated and the correlated case, it is necessary to interfere the state with an independent reference beam. This can be understood as a purity measurement of the marginal beams.

\begin{figure}[ht]
\begin{center}
\includegraphics*{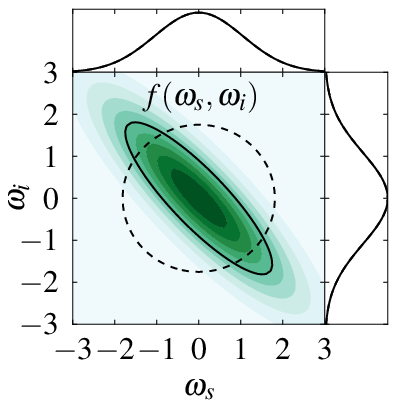} \quad \includegraphics*{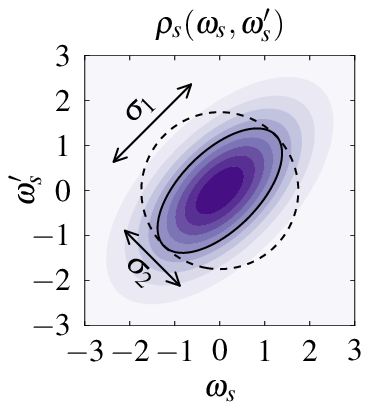} \quad \includegraphics*{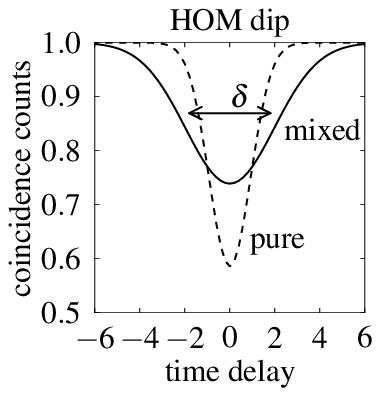}
\caption{Comparison of an uncorrelated PDC state (dotted black lines) to a correlated PDC state (green, purple or solid black lines) . Left: JSA. Center: reduced density matrix of signal. Right: HOM dip between signal and reference. Both cases, correlated and uncorrelated, have the same marginal spectral widths. Nevertheless, the width of the HOM dip reveals the amount of correlations.} 
\label{fig:theory_purity}
\end{center}
\end{figure}

If signal and idler are correlated, tracing out the idler mode results in a mixed signal state, described in frequency space by a density matrix $\rho_s(\omega_s, \omega'_s)$. Being hermitian, this matrix must be symmetric with respect to the principal diagonal. In the following we assume that it is a 2-dimensional Gaussian function, which is well justified for our source as discussed in Sec. \ref{sec:spectral_measurements}. It follows from straight forward calculations that the purity of the state is given by $\textrm{Tr} (\rho_s^2) = \sigma_2/\sigma_1$,  where $\sigma_1$ and $\sigma_2$ are  the major and minor axes of the Gaussian function $\rho_s(\omega_s, \omega'_s)$. The major axis is the spectral width of the signal beam and can be measured directly with a spectrometer. The minor axis cannot be accessed directly with a spectral measurement but has an influence on the coherence length of the state and hence the interference pattern, as depicted in Fig. \ref{fig:theory_purity}. By a HOM interference measurement with a known reference field, the minor axis can be deduced from the temporal width of the HOM dip, which is given by
\be \delta^2 = \frac{1}{2\sigma_2^2} + \frac{1}{2\sigma_\beta^2} \label{eq:HOM_width}\ee
where $\sigma_\beta$ is the width of the reference field. A detailed analysis of this technique can be found in \cite{cassemiro_accessing_2010}. Note that a measurement of the widths is much more robust  compared to a measurement of the visibility because it is not affected by higher photon number components and imperfect  overlaps of the state in the spectral and polarization degree of freedom. For comparison with the JSA, the relation between the purity and the Schmidt number is simply given by $P = 1/K$.

\paragraph{Joint spectral intensity and second order correlation function}
Alternatively to HOM interference measurements, information about the separability of the state can be inferred from the joint spectral intensity (JSI) $|f(\omega_s, \omega_i)|^2$ or the second order Glauber correlation function $g^{(2)}(\tau)$ \cite{glauber_quantum_1963} of the signal or idler field without heralding on a photon in the opposite mode. The $g^{(2)}(0)$ value gives information about the photon number statistics of the marginal beams. For a decorrelated PDC state, the marginal statistics are thermal with $g^{(2)}(0)=2$ and for a strongly correlated state, the marginal statistics are Poissonian with $g^{(2)}(0)=1$ \cite{tapster_photon_1998}. The relation between the Schmidt number and the marginal Glauber correlation function is given by $g^{(2)}(0)=1+1/K$ \cite{christ_probing_2011, eckstein_highly_2011}. 

All three methods for source characterization have already been applied in the literature. However, there is no ideal measurement: A JSI measurement is affected by a low spectral resolution and cannot resolve correlations hidden in the phase of the JSA.  Furthermore, it is blind to any non spectral correlations. A $g^{(2)}(0)$ measurement is sensitive to correlations in all degrees of freedom, but is, unfortunately, strongly affected by detector dark counts and non-PDC background signal. The HOM interference measurement requires the exact knowledge of spectral widths of signal and reference fields. Therefore it makes sense to apply all three techniques and carefully compare their results as we do in this paper.

\section{Experimental results}
\label{sec:experiment}
\subsection{Experimental setup}

We employ type-II PDC in ppKTP waveguides. At telecom wavelength it offers a positively correlated phasematching function with a correlation angle of $59$ degrees, such that a decorrelated JSA can be produced using the correct pump width \cite{eckstein_highly_2011}. Such waveguides can be routinely produced and are commercially available. Our chip is purchased from ADVR and has the following parameters: a length of $\unit[8]{mm}$, waveguide dimensions of about $\unit[4]{\mu m} \times\unit[6]{\mu m}$ and a poling period of $\unit[117]{\mu m}$.  

\begin{figure}[ht]
\begin{center}
\includegraphics[width=0.9\linewidth]{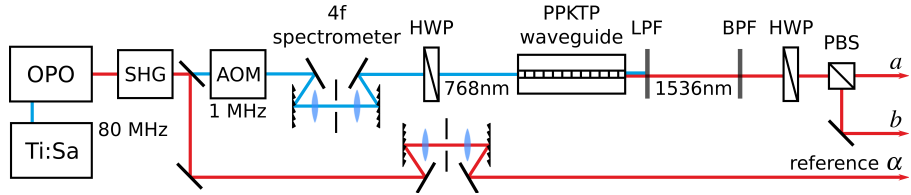}
\caption{Scheme of the setup. Pulsed light at telecom wavelengths is generated in an Ti:Sapph pumped optical parametric oscillator (OPO) and frequency doubled by second harmonic generation (SHG). Part of the light is separated for later use as a reference field. The repetition rate is lowered by an acousto-optical modulator (AOM) to match the repetition rate of the gated APDs. The 4-f spectrometer tailors the spectral width of the pump beam to achieve spectral decorrelation. The PDC state is generated inside the periodically poled KTP waveguide. The pump is separated by a long pass filter (LPF). A bandpass filter (BPF) is used to suppress background outside the PDC spectrum. Finally signal and idler are separated at a polarizing beam splitter (PBS).}
\label{fig:setup}
\end{center}
\end{figure}

A scheme of the setup is shown in Fig. \ref{fig:setup}. It consists of a Ti:Sapph pumped OPO (Chameleon Compact OPO) running at $\unit[80]{MHz}$ and producing short pulses of $\unit[250]{fs}$ at $\unit[1536]{nm}$. A small part of the light is separated for the reference beam while the major part is frequency doubled in a BBO crystal. The cascade of Ti:Sapph, OPO and SHG is necessary to achieve a good synchronization between the coherent reference beam and the PDC light. A crucial part of the experiment is the spectral shaping of the pump. We employ a $4f$-spectrometer, consisting of two gratings, two lenses and one slit in the center, all separated by the focal length of the lenses. The width of the spectrum can be narrowed by reducing the width of the slit. Additionally, the slit is tilted, effectively reducing the resolution of the spectrometer, to produce Gaussian spectral shapes rather than square-like shapes. The spectrum of both, pump and reference beams, is measured and monitored by an optical spectrum analyzer (OSA). 

We use aspheric lenses for in- and outcoupling of the waveguide and the fibers. Behind the waveguide, we employ a longpass filter (Semrock LP02-808RU-25) to block out the pump beam and a bandpass filter (Semrock NIR01-1535/3-25) with an FWHM of $\unit[8]{nm}$. This bandpass filter is the second crucial part of the experiment because the PDC generates background signal over a wide spectral range \cite{eckstein_realistic_2011}. The bandpass filter is chosen such that it blocks as much background signal as possible without affecting the PDC signal itself. All beams are coupled into single mode fibers (SMF). For the detection at the single photon level, we use avalanche photo diodes (APDs) (either Id Quantique id201 or NuCrypt CPDS-1000-4). The repetition rate of the id201 is \unit[1]{MHz} and of the CPDS-1000-4 \unit[40]{MHz}. We measure raw coincidence vs single ratios, i.e. Klyshko efficiencies \cite{klyshko_utilization_1977, rarity_absolute_1987, achilles_direct_2006} of up to $20.5\%$ for the signal beam and $15.5\%$ for the idler beam. Corrected for the detector (id201) efficiencies of approximately $25\%$ and $22\%$, these values correspond to coupling efficiencies into the SMFs of roughly $80\%$ in one arm and $70\%$ in the other arm, making our source highly compatible with existing fiber based telecommunication technologies.

\begin{figure}[ht]
\begin{center}
\includegraphics[width=0.98\linewidth]{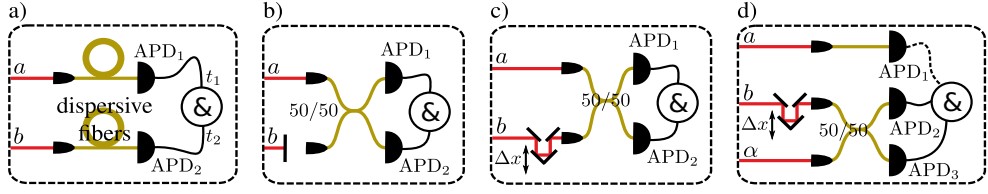}
\caption{Scheme of different measurement settings. a) JSI measurement; b) $g^{(2)}(0)$ measurement; c) HOM interference between signal and idler; d) 2-fold or 3-fold HOM interference with a reference field.  The Id Quantique detectors($\unit[1]{MHz}$) are used for the JSI measurement and the NuCrypt detectors ($\unit[40]{MHz}$) for the rest.}
\label{fig:setup2}
\end{center}
\end{figure}

\subsection{Spectral measurements}
\label{sec:spectral_measurements}
We characterize the spectral properties of our source with a fiber spectrometer \cite{avenhaus_fiber-assisted_2009} (see first frame of Fig. \ref{fig:setup2}). Signal and idler travel through dispersive fibers before impinging onto two APDs. Different wavelengths arrive at different times. By scanning the gating times of the APDs, a joint spectral intensity (JSI) distribution is obtained from coincidence click rates. Similarly, the marginal spectral distribution of signal and idler are obtained from the single click rates. The resolution of the fiber spectrometer is limited by the gate width of the APDs and the length of the fiber. We use the id201 with a gate width of approximately \unit[1.5]{ns}, resulting in a spectral resolution of \unit[1.8]{nm} for the JSI and \unit[0.9]{nm} for the marginal spectra.

\begin{figure}[ht]
\begin{center}
\includegraphics*{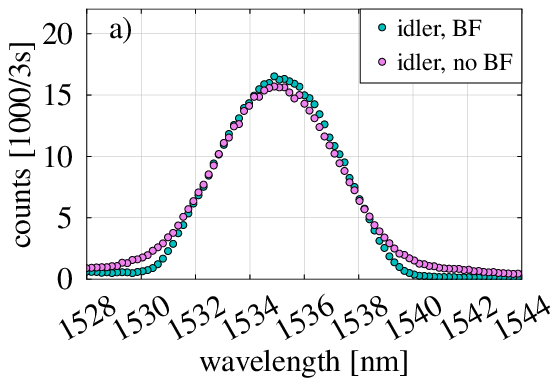} \qquad \includegraphics*{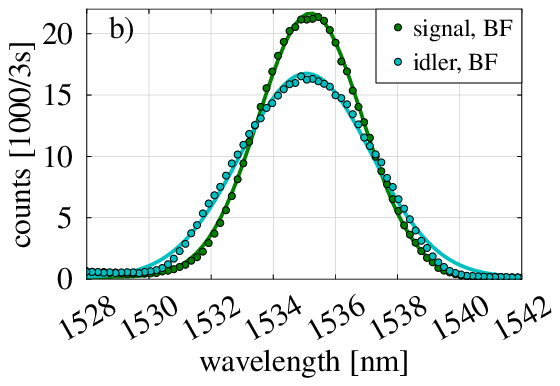} \\ \includegraphics*{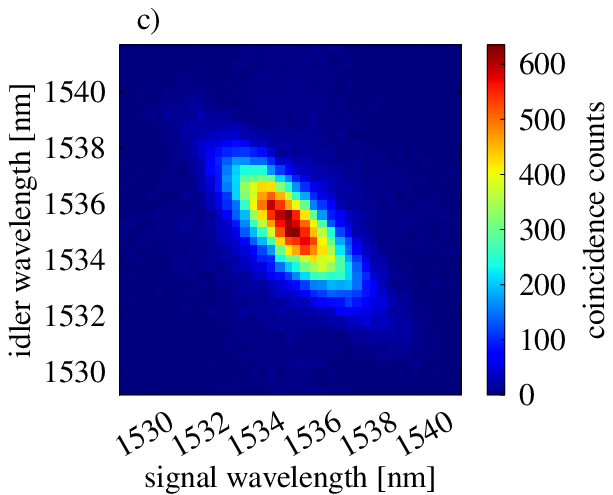} \quad \includegraphics*{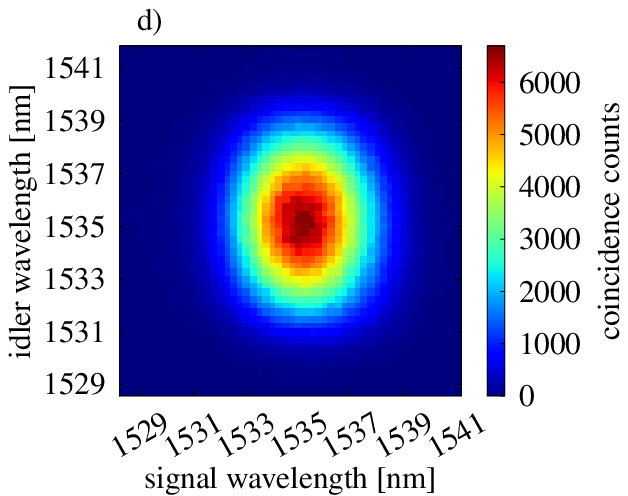}
\caption{a) Marginal idler spectra for the decorrelated case with and without the bandpass filter (BF). One can see that the spectral filter fits nicely the idler spectrum, touching it only slightly at the very edges. b) Signal and idler spectra with the bandpass filter. Bottom: Measured joint spectral intensity for anticorrelation (c) and decorrelation (d). The pump widths are $\unit[0.5]{nm}$ and $\unit[2.1]{nm}$, respectively. }
\label{fig:JSI}
\end{center}
\end{figure}

The JSI and the marginal spectral distributions with and without the bandpass filter are shown in fig. \ref{fig:JSI}. The bandpass filter has a width of \unit[8]{nm} and one can see from the comparison of the unfiltered with the filtered spectrum in Fig. \ref{fig:JSI} that background signal is suppressed directly outside of the PDC range while the PDC spectrum itself is mostly unaffected. The spectra of signal and idler have Gaussian shapes and their widths at FWHM, obtained from Gaussian fits, are \unit[5.2]{nm} and \unit[4.0]{nm}, respectively. Assuming that the finite resolution is effectively a convolution with an \unit[0.9]{nm} wide Gaussian, the true widths can be estimated to be \unit[5.1]{nm} and \unit[3.9]{nm}. The theoretical sinc-shape of the spectrum is smeared out by experimental imperfections of the waveguide structure. The congruence of the data points with the fit curve supports the assumption of a Gaussian JSA in Sec. \ref{sec:HOM_interference}.

The JSI measurement in the decorrelated case shows a smooth, nearly round shape. This also supports the assumption of Gaussian phasematching and pump spectral profiles. Being an intensity distribution, the JSI does not provide full information about the state as additional phase or temporal information would be required to reconstruct the JSA \cite{brecht_characterizing_2013}. A singular value decomposition of the square root of the JSI, yielding an effective mode number of $1.001$ in our case, can therefore be only regarded as a lower bound for the true Schmidt number. 

To obtain a stronger statement about the correlations present in the source, we measure the second order Glauber correlation function $g^{(2)}(0)$ with a $50/50$ fiber coupler \cite{tapster_photon_1998}, as depicted in the second frame of Fig. \ref{fig:setup2}. We find raw values of $1.83\pm0.02$ and $1.86\pm0.02$ for signal and idler, respectively. These values pose an upper bound on the correlations present in the PDC source. The corresponding Schmidt numbers are $1.20$ and $1.16$. Part of the deviation from the perfectly decorrelated case could be caused by background events. From comparison of the filtered with the unfiltered spectra, we conclude that the background still remaining under the PDC spectrum makes up for approximately $\unit[1.9]{\%}$ of the total count rates. Since background signal has a Poissonian photon number distribution, it strongly degrades the $g^{(2)}(0)$ value. Correcting for these background events \cite{eckstein_realistic_2011}, we get values of $1.90$ and $1.94$. However, we would like to note that the raw value of $g^{(2)}(0)=1.86$ is among the highest compared to other PDC sources \cite{tapster_photon_1998, eckstein_highly_2011, li_fiber-based_2008, takesue_entanglement_2009} indicating that the amount of background events is relatively low.

\subsection{Interference measurements}
\begin{figure}[ht]
\begin{center}
\includegraphics*{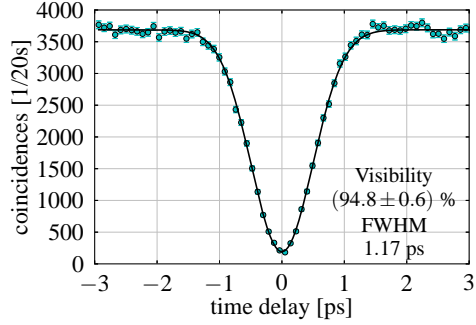}
\caption{HOM interference between signal and idler.}
\label{fig:dip1}
\end{center}
\end{figure}

To demonstrate the indistinguishability between signal and idler, we interfere them at a $50/50$ fiber coupler, as sketched in the third frame of Fig. \ref{fig:setup2}. The coincidence rate versus a delay of one of the beams is shown in Fig. \ref{fig:dip1}. The measurement is done with a pump energy as low as $\unit[0.6]{pJ}$ per pulse leading to a mean photon number of $0.002$. We obtain a visibility of $(94.8\pm 0.6)\%$, which is to our knowledge the highest value reported in the literature for PDC sources without narrowband filtering. It is close to the theoretical value of $96.7\%$ for the decorrelated case. Part of that small deviation from the theoretical value is caused by our fiber coupler which has a slightly uneven coupling ratio of $49.1/50.9$. The high visibility of the interference shows that the indistinguishability between signal and idler is indeed very high.

\begin{figure}[ht]
\begin{center}
\includegraphics*{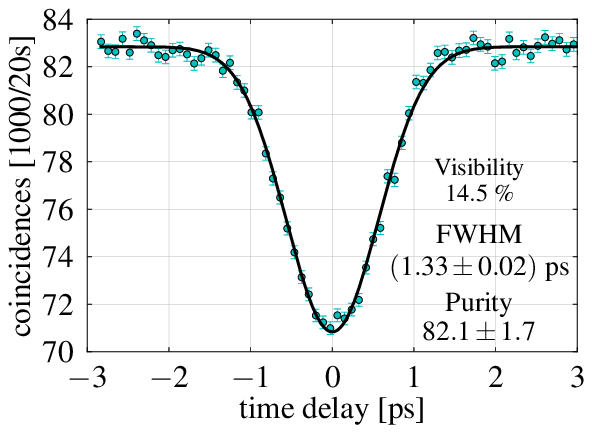} \qquad \includegraphics*{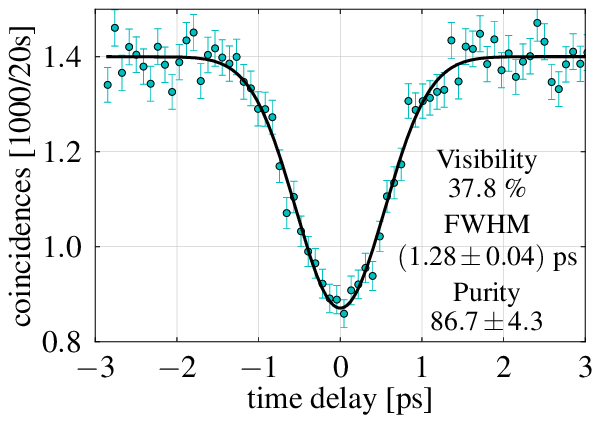}
\caption{two-fold (left) and three-fold (right) HOM interference between signal and reference.}
\label{fig:dip2}
\end{center}
\end{figure}

As discussed in the theory section, a robust method for verifying the decorrelation of the PDC state is to measure the purity of the marginal beams by HOM interference with a reference field. As the reference field, we use part of the original laser beam and attenuate it to the single photon level. All three beams, signal, idler and reference are coupled into single mode fibers and the count rates are recorded with the APDs, as sketched in Fig. \ref{fig:setup2}. We record two-fold and three-fold coincidences, where by two-fold we mean coincidence events behind the beam splitter disregarding the third APD and by three-fold we mean triple coincidences between all three output ports. The heralding with the second PDC beam in the three-fold case increases the visibility of the interference \cite{laiho_producing_2009}. The results are shown in Fig. \ref{fig:dip2}. As expected, the visibility in the three-fold case is higher than in the two-fold case. It depends on the spectral overlap between the two beams as well as the mean photon numbers which for this measurement are $0.006$ for the marginal beam and $0.08$ for the reference beam. For calculating the purity of the state, we use the width of the dip rather than the visibility. Both two-fold and three-fold curves have similar widths of $\unit[1.33\pm0.02]{ps}$ and $\unit[1.28\pm0.04]{ps}$. Taking into account the spectral width of the Gaussian reference field of $\unit[4.5]{nm}$ and the signal spectral width of $\unit[3.9]{nm}$, we calculate a purity of $\unit[82.1 \pm 1.7]{\%}$ and $\unit[86.7 \pm 4.3]{\%}$. These values are in good agreement with the raw $g^{(2)}(0)$ values. Compared to other PDC sources without narrowband filtering, Mosley et. al \cite{mosley_heralded_2008} have shown higher purities around $\unit[95]{\%}$ by measuring the HOM interference between two independent PDC sources. Our measured purities are slightly below this value but still in the same range, demonstrating excellent source performance. Similar values have been obtained in \cite{jin_high-visibility_2011}, who also utilized HOM interference with a coherent field. 

\subsection{Source brightness}
In the low power regime at which the above interference measurements are performed, we measure a photon pair generation efficiency of $\unit[3\times 10^9]{\textrm{(photon pairs)}/J}$ or $\unit[3.8\times 10^{11}]{\textrm{(photon pairs)}/(J\times m)}$, if taken per unit crystal length and pump energy. Taking into account that all photons are generated in a single mode, the source brightness is extremely high, even for low pump powers. For a mean  photon number per pulse of $0.1$ in the signal beam, the required cw equivalent power at a repetition rate of $\unit[1]{MHz}$ is only $\unit[33]{\mu W}$. By using the full available power in our setup of $\unit[2.5]{mW}$ ($\unit[2.5]{nJ}$ per pulse), we are able to reach a mean photon number of approximately $80$ photon pairs per pulse. This high number is due to the single mode character of our source and much higher than is expected for a multimode source with the same photon pair generation efficiency.

\section{Conclusion}
\label{sec:conclusion}
In conclusion, we have implemented a source  with remarkable properties in terms of brightness, purity and symmetry. The HOM interference between signal and idler showed, to our knowledge, an unprecedented visibility of $94.8\%$. The purity values for signal and idler obtained from $g^{(2)}$, JSI and interference with a reference beam reveal a purity of above $\unit[80]{\%}$. In order to reach a desirable mean photon number around $0.1$ photon pairs per pulse, pump energies as low as $\unit[33]{pJ}$ are required, which significantly simplifies the simultaneous operation of several sources. Our source can be easily combined with identical sources or other optical fields and constitutes a step towards the feasibility of more complex quantum circuits or networks. \\

\section*{Acknowledgement}
We thank Kaisa Laiho, Andreas Eckstein and Malte Avenhaus for valuable and helpful discussions. The research leading to these results has received funding from the European Community's Seventh Framework Programme FP7/2001-2013 under grant agreement no. 248095 through the Integrated Project Q-ESSENCE.

\end{document}